\documentclass{ws-p8-50x6-00}

\def\today{\it November 3, 2000}

\begin{document}

\def\Ds{D\!\!\!\!/\,}
\def\qt{{\tilde{q}}}
\def\qbar{{\overline{q}}}
\def\qtbar{{\overline{\qt}}}

\title{Chiral Perturbation Theory, Non-leptonic Kaon Decays, and the Lattice}

\author{Maarten Golterman}

\address{Department of Physics, Washington University, St. Louis MO 63130,
USA\\E-mail: maarten@aapje.wustl.edu}


\maketitle

\abstracts{
In this talk, I first motivate the use of Chiral Perturbation Theory
in the context of Lattice QCD.  In particular, I explain how
partially quenched QCD, which has, in general, unequal valence- and 
sea-quark masses, can be used to obtain 
real-world ({\it i.e.} unquenched) results for low-energy
constants.  In the second part, I review how Chiral Perturbation
Theory may be used to overcome theoretical difficulties which
afflict the computation of non-leptonic kaon decay rates from
Lattice QCD.  I argue that it should be possible to determine
at least the $O(p^2)$ weak low-energy constants reliably
from numerical computations of the $K\to\pi$ and $K\to$\ vacuum
matrix elements of the corresponding weak operators.}

\section{Introduction}
Ideally, Lattice QCD should lead to the unambiguous, first-principle
determination of hadronic quantities.  However,
it does not work that way in practice, basically because of 
limits on available computational power.  Because of these limits,
approximations are needed which lead to systematic errors, 
and we need a theoretical
understanding of these in order to obtain 
estimates of hadronic quantities with known and controlled errors.
This talk addresses some of the limitations of Lattice QCD
with respect to the physics of Goldstone Bosons (GBs), and
explains how Chiral Perturbation Theory (ChPT) can be used to extract
physical quantities from the often unphysical results of lattice
computations.  

The basic idea is that ChPT parametrizes QCD correlation functions
involving GBs in terms of a number of parameters, the low-energy
constants (LECs).  To a given order in the chiral expansion,
physical quantities are determined completely in terms of a
finite number of LECs, and these LECs can then be obtained from
(possibly other) correlation functions at often unphysical choices
of masses and momenta which are more easily accessible
to numerical computation.  Such a program only
works if the chiral expansion converges in the regions of
interest (both physical and on the lattice), but this can in
principle be checked by considering successive terms in the
chiral expansion, once the LECs have been determined from fits to
lattice results.  Obviously, for such checks at least two orders
in the chiral expansion need to be considered, and one-loop
ChPT is essential.

A simple example is the pion mass, which is extracted from
the appropriate euclidean two-point function.  The computations
are done in a finite volume $L^3$, and, to avoid finite-%
size distortions, one has to require that $m_\pi L\gg 1$.  Since,
in practice, $L\approx 3$~fm is (currently) the largest attainable
volume, demanding (say) $m_\pi L>5$ leads to the requirement
that $m_\pi>330$~MeV on the lattice; in other words, the
light quark masses have to be chosen unphysically large.  But if
ChPT converges in the range $300$~MeV$<m_\pi
<700$~MeV or so, it can be used to determine the relevant LECs
and extrapolate to the physical pion mass ({\it e.g.} 
to determine light quark masses).  
Once this works, there is actually no need
to perform numerical computations in volumes large enough
to contain a physical pion.  Instead, additional computational
power could for instance be used to go to smaller lattice spacings
at the same physical volume, in order to reduce scaling violations.

In this talk, I will discuss two topics from this point of view. 
The first topic will be the use of ``partially quenched" QCD
(PQQCD), which extends the idea that, for a number of quantities
of interest, simpler unphysical lattice computations may be all
one needs.  The second topic will be that of weak matrix elements
for non-leptonic kaon decays.  In this case, accessible
euclidean correlation functions are ``not physical" for a number
of reasons, but ChPT can be used to extract relevant information. 

\section{Partially Quenched QCD}
\subsection{Why partially quenched?}
In a lattice calculation, the quark fields are integrated over
analytically, leading to a gluonic expectation value of the
product of the fermion determinant and various quark propagators.  
For example,
\begin{equation}
\langle{\overline u}\gamma_5 d(x){\overline d}\gamma_5 u(0)\rangle
\!=\!\frac{\langle\gamma_5(\Ds+m_d)^{-1}(x,0)\gamma_5(\Ds+m_u)^{-1}(0,x)
{\rm det}(\Ds+m)\rangle_{\rm glue}}
{\langle{\rm det}(\Ds+m)\rangle_{\rm glue}}.\label{PIONCORR}
\end{equation}
It is clear that, in a computation, one could choose the quark mass
in the determinant (the ``sea" quark mass) unequal to those in the 
propagators (the ``valence" quark masses), as indicated in 
eq.~(\ref{PIONCORR}). This situation is referred to as ``partial
quenching."  The (fully) quenched theory is obtained by omitting the
sea quarks altogether, or, equivalently, by sending $m=m_{\rm sea}\to\infty$.
A simple reason to do this is that it is much cheaper to compute
a propagator than a determinant, so that one can vary the valence masses
almost ``for free," unlike the sea-quark mass.  This then raises two
questions: is this useful for anything, and, can this be understood with
field-theoretic methods?

To the first question, the answer is yes!  For example,
a partially quenched simulation with $N=3$ degenerate light sea quarks (and
arbitrary valence masses) gives information about the real 
world.\cite{shasho}  
Consider again the pion mass.  With $N$
sea quarks of mass $m_s$, and two valence quarks
of mass $m_v$, partially quenched ChPT (PQChPT; see next subsection) 
gives\cite{sharpe,golleu}
\begin{eqnarray}
m_\pi^2&=&2B(N)m_v\Biggl(1+{\rm chiral\ logs} \label{PIONMASS} \\
&&+\frac{32B(N)}{f^2(N)}\Bigl[\left(2L_8(N)-L_5(N)\right)m_v+
\left((2L_6(N)-L_4(N)\right)Nm_s\Bigr]
\Biggr)\,,\nonumber
\end{eqnarray}
where we follow the notation of Gasser and Leutwyler\cite{gasleu} (but the pion 
decay constant $f(N)$ in the chiral limit is normalized such that it equals
$132$~MeV at $m_\pi$).  We explicitly indicate the dependence on 
$N$.  (The QCD dynamics ({\it i.e.} the determinant) and hence the LECs
do not depend on the valence quarks.)
Since the LECs depend {\it only} on $N$, it is clear that if we take $N=3$,
the values obtained from fitting this expression to lattice results are the
real-world values.  Note that only degenerate sea quarks are needed to get
$2L_8-L_5$ and $2L_6-L_4$ separately.  If we only use unquenched QCD,
non-degenerate sea quarks would be needed to get the same information,
or one would have to consider matrix elements like $\langle\pi|{\overline s}
s|\pi\rangle$, which involve -- numerically much harder --
disconnected diagrams.  It is clear that the $N=3$ LECs cannot be obtained
from quenched QCD, which has $N=0$.
If we can simulate PQQCD with three dynamical degenerate quarks at roughly
the strange quark mass to a high enough accuracy to fit eq.~(\ref{PIONMASS}),
one could thus, for example, decide the fate of the Kaplan--Manohar
ambiguity\cite{kapman} from first principles.

\subsection{Partially quenched field theory}
PQQCD can be defined as a path integral, by considering the 
lagrangian\cite{bgpq}
\begin{equation}
{\cal L}=\qbar^i_v(\Ds+m_v^i)q^i_v+\qbar^\alpha_s(\Ds+m_s^\alpha)q^\alpha_s
+\qtbar^j(\Ds+m_v^j)\qt^j\,, \label{PQQCDL}
\end{equation}
where $q^i_v$, $i=1\dots K$ are the valence quarks and $q^\alpha_s$,
$\alpha=1\dots N$ the sea quarks.  The valence quarks are made into such
by adding $K$ ``ghost" quarks $\qt^j$, $j=1\dots K$, which are identical to
the valence quarks $q_v^i$ (with the same masses $m_v^i$), except that 
they are given opposite statistics.\cite{morel}  Therefore, the valence
and ghost determinants cancel, leaving only the sea-quark determinant,
as in eq. (\ref{PIONCORR}).  Obviously, the theory defined in this way
is not a ``healthy" field theory (it violates spin-statistics), but the
interesting point is that one can study its unphysical, euclidean
correlation functions, and extract information about the real world,
as in the example of the previous subsection.
 
For this, we need to adapt ChPT to the partially quenched situation.
This can be done systematically by observing that ${\cal L}$ has a
large chiral symmetry group, $SU(N+K|K)_L\times SU(N+K|K)_R$,
where $SU(N+K|K)$ is a graded version of $SU(N+2K)$ because it 
transforms fermions ($q_v$, $q_s$) into bosons ($\qt$), and
{\it vice versa}.  Based on this, PQChPT can be
developed in much the same way as in the usual case, using
supertraces and superdeterminants to build graded-group
invariants.\cite{bgq,bgpq}  Quenched QCD corresponds to 
$N=0$.  Note that partially quenched QCD, unlike quenched QCD,
contains QCD (for which $N=3$, $K=0$) as a special case.%
\cite{bgpq}
 
I have not enough space to review (P)QChPT,\cite{golzak} 
and will limit myself to a brief discussion
of neutral mesons ({\it i.e.} those containing
quarks and anti-quarks of the same flavor).  The tree-level
two-point function for mesons $\qbar^i\gamma_5 q^i$ (no sum
over $i$) is,\cite{bgpq,sharpe,shasho,golpal} in the limit of large
$\eta'$ mass, equal to
\begin{equation}
\delta_{ij}\frac{1}{p^2+M_{VV}^2}-\frac{1}{N}\frac{1}{p^2+M_{VV}^2}
-\frac{1}{N}\frac{M_{SS}^2-M_{VV}^2}{(p^2+M_{VV}^2)^2}\,,
\label{NEUPROP}
\end{equation} 
where $M_{VV}$ ($M_{SS}$) is the mass of a meson made from (degenerate)
valence (sea) quarks.  In the quenched theory ($N=0$),
the $\eta'$ cannot be decoupled, leading
to a two-point function\cite{bgq,sharpe2}
\begin{equation}
\delta_{ij}\frac{1}{p^2+M_{VV}^2}-\frac{m_0^2/3}{(p^2+M_{VV}^2)^2}\,,
\label{QNEUPROP}
\end{equation}
where $m_0$ is the parameter corresponding to the singlet part of
the $\eta'$ mass in the unquenched theory.
The remarkable new feature in these two-point functions is the double-pole
term.  (Note that it disappears for $M_{VV}=M_{SS}$, $N\ne 0$,
as it should.)  This double pole leads to several infrared diseases,
which are a consequence of (partial) quenching, and not present in
full QCD:
\begin{itemize}
\item ``Enhanced chiral logarithms."\cite{bgq,bgpq,sharpe2}  Instead of
having the usual chiral logs like $M^2\log{M^2}$, one finds logs such as
$(M_{SS}^2-M_{VV}^2)\log{M_{VV}^2}$ ($N\ne 0$) or $m_0^2\log{M_{VV}^2}$
($N=0$), which diverge in the chiral limit $M_{VV}\to 0$.
\item ``Enhanced finite-volume effects."\cite{bgpp,golpal99}  See subsection
3.2.
\end{itemize}

\subsection{Quenched ChPT and Lattice QCD}
An important question is whether numerical computations are precise
enough to discern one-loop effects in ChPT.  The most obvious place to
look is the quenched numerical results for GB masses and decay constants,
for which the most extensive and statistically significant numerical
results are available.  In quenched ChPT, one finds,\cite{bgq,sharpe2}
for degenerate quarks
\begin{eqnarray}
\frac{m_\pi^2}{8\pi^2 f^2}&=&y\left(1-(\delta-\frac{2}{3}\alpha y)
\log{y}-\delta+\frac{1}{3}\alpha y
+8(2\lambda_8-\lambda_5)y\right)\,, \label{PIONQUENCHED} \\
f_\pi/f&=&1+4\lambda_5 y\,, \nonumber
\end{eqnarray}
where $\lambda_i=(4\pi)^2 L_i$ are the (quenched) $O(p^4)$ LECs, and
\begin{equation}
y\equiv\frac{2Bm}{8\pi^2 f^2}\,,\ \ \ \delta\equiv\frac{m_0^2/3}{8\pi^2 f^2}\,.
\label{ABBREV}
\end{equation}
(There are similar expressions for non-degenerate quark masses.)
The parameter $\alpha$ parametrizes momentum dependence of the singlet
GB two-point vertex, and appears in the renormalization of the $\eta'$ 
field.\cite{bgq,gasleu}
There are really two questions: 1) Given certain numerical results, can we
see the chiral logs? and 2) Can we determine $\lambda_{5,8}$?  

CP-PACS\cite{cppacs} has high statistics results at quark masses ranging
from $m_\pi\sim 300$ to $750$~MeV, extrapolated to the
continuum limit, and shows that this is
sufficient to fit $\delta$ from eq.~(\ref{PIONQUENCHED}), setting
$\alpha=0$.  With this restriction, they see the chiral logs,
and find $\delta\approx 0.1$.  This answers the first question.  A full
analysis of their results has not appeared yet, but 
estimates for $\lambda_{5,8}$ can also be extracted.  From their fits,
I find, for example, $10^3(2L_8-L_5)\approx 0.3$.\cite{yoshie}
With these values, eq.~(\ref{PIONQUENCHED}) appears to converge well.
These numbers were recently confirmed by Duncan {\it et al.},\cite{dunetal} who
find $\delta=0.065(13)$ (at one value of the lattice spacing, with 
improved Wilson fermions), $10^3L_5=2.5(5)$ ({\it cf.} the value of the
Alpha collaboration below!) and $10^3(2L_8-
L_5)=0.4(2)$ (again, taking $\alpha=0$).  It remains to be
seen what can be said when $\alpha$ is not constrained to vanish.

The Alpha collaboration\cite{alpha} has recently presented high-statistics,
continuum-extrapolated results in the range $590-670$~MeV, which,
as they point out, is not sufficient to demonstrate the presence of
chiral logs.  However, using eq.~(\ref{PIONQUENCHED}) with the CP-PACS
value of $\delta=0.12(2)$ (assuming that the one-loop chiral expansion
converges for this mass range), they find, for $\alpha=0$
\begin{equation}
10^3L_5=0.78\pm 0.05\pm 0.20\,,\ \ \ 10^3(2L_8-L_5)=0.28\pm 0.04
\pm 0.10\pm 0.06\,,\label{ALPHAONE}
\end{equation}
where the 1st (2nd, 3rd) error is statistical (higher-order uncertainty,
error on $\delta$).  Arguing that, if $\alpha=0.5$, $\delta=0.05(2)$
would be reasonable, they find
\begin{equation}
10^3(2L_8-L_5)=0.02\,,\label{ALPHATWO}
\end{equation}
with similar errors.  It is clear that this value is very sensitive to
$\alpha$.  All these estimates are quenched, and should not
be directly compared to the real world.  (The quenched LECs considered
here do not even run.)  It should also be noted that in partially
quenched simulations, if the quark masses are small enough, the $\eta'$
decouples, and the parameters $\delta$ and $\alpha$ do not appear in 
the ChPT expressions\cite{shazha} 
--- the coefficients in the logs are fully determined
by chiral symmetry.  The logarithm in eq.~(\ref{PIONQUENCHED}) is a
quenching artifact.  (However, $\delta$ and $\alpha$ 
do appear in PQChPT if the quark masses are
not small enough.\cite{golleu})

\section{Non-leptonic two-body kaon decays}
\subsection{$\Delta I=3/2$}
The simplest non-leptonic kaon decay is
$K^+\to\pi^+\pi^0$, which is pure $\Delta I=3/2$.  In order to determine this
decay rate, one needs the weak matrix element
\begin{equation}
\langle\pi^+\pi^0|{\overline s}_L\gamma_\mu
(d_L{\overline u}_L\gamma_\mu u_L-u_L{\overline d}_L\gamma_\mu d_L
+u_L{\overline u}_L\gamma_\mu d_L)|K^+\rangle\,,
\end{equation}
where $u_L$ denotes the left-handed part of $u$, {\it etc}.  On the lattice,
one computes euclidean correlation functions, and this matrix
element is contained in
\begin{equation}
C(t_2,t_1)=\langle 0|\pi^+(t_2)\pi^0(t_2)O_{\rm weak}(t_1)K^-(0)|0\rangle\,,
\label{CORR}
\end{equation}
where $\pi^0(t_2)$ is a pion field with arbitrary momentum ${\vec q}$
at time $t_2$, {\it etc.} (we take the kaon to be at rest). Inserting
complete sets of states on both sides of $O_{\rm weak}$, and keeping only 
the leading exponential in $t_1$ on the kaon side, we find (with
$t_2>t_1>0$)
\begin{equation}
C(t_2,t_1)\sim\sum_n\langle\pi^+(-{\vec q})\pi^0({\vec q})|n\rangle
\langle n|O_{\rm weak}|K^+\rangle\langle K^+|K^-|0\rangle
e^{-E_n(t_2-t_1)-m_K t_1}\,. 
\label{CORRAPPROX}
\end{equation}
The leading exponential corresponds to the state
$|n\rangle$ with the lowest energy, and this is the state with both 
pions at rest ({\it i.e.} ${\vec q}=0$), with $E=2m_\pi$!
This is a manifestation of the so-called Maiani--Testa theorem,\cite{maites}
which states that it is impossible to obtain the physical matrix 
element from the large-time behavior of the euclidean correlation
function when there are more particles in the final state.  
All we get is the {\it unphysical} matrix element with
all mesons at rest.  Note that this conclusion does not depend on
what we pick ${\vec q}$ to be: only the total momentum in the decay
is conserved. The desired matrix element, for which $|{\vec q}|=
\frac{1}{2}\sqrt{m_K^2-4m_\pi^2}$, is buried in the excited states. 
The insertion of $O_{\rm weak}$ at a fixed time does not conserve energy. 

At present, there are two approaches to this problem.  One approach
is to make use of the fact that the desired physical matrix element
can be obtained from $C(t_2,t_1)$ in a finite spatial
volume $L^3$, in which the excited energy levels are discrete.\cite{lellue}
With periodic boundary conditions, the energies are quantized
roughly as $E_{\vec n}=2\sqrt{m_\pi^2+(2\pi{\vec n})^2/L^2}$ (plus $O(1/L^3)$
corrections due to final-state interactions (FSI)), and 
one finds that the first excited states $|{\vec n}|=1$ have energy
equal to the incoming energy $m_K$ if $m_\pi L\approx 4$ for realistic
meson masses.  So, if it is possible to determine the first
excited-state exponential in $C(t_2,t_1)$ reliably on the lattice, one obtains
the physical matrix element in a finite (but rather large) volume.
This matrix element can be then converted to infinite volume, with a 
correction factor rather close to one.\cite{lellue}

This method, while theoretically ``clean," may turn out to be
hard to implement in practice.  First, one needs numerical computations
with a high enough precision to extract the first excited state.
Then, one needs $2m_\pi<m_K<4m_\pi$ (the latter
inequality such that the rescattering of the pions is elastic), and
this implies rather small up and down
quark masses (of order $m_{\rm strange}/8$ or less) if $m_K\approx 500$~MeV.
For light 
quark masses at $\sim m_{\rm strange}/8$ this means
$L\approx 5$~fermi, which is large by present standards.  
With a larger lattice kaon mass, $L$ can be smaller.
ChPT can be used to extrapolate to physical masses.\cite{bpp}
Because of the difficulty of the required numerical computations,
one would like to test these ideas in a (partially) quenched
setting, but it is not clear at the moment what the effects of
quenching would be in this approach.  (For instance, ``enhanced"
finite volume effects may occur,\cite{bgpp,golpal99} see below.)
It would be interesting to investigate this in (P)QChPT. 

An alternative, and in fact older, idea is to compute the unphysical
matrix element with all mesons at rest from the dominant term in 
eq.~(\ref{CORRAPPROX}), for which 
$|n\rangle=|\pi^+({\vec 0})\pi^0({\vec 0})\rangle$,\cite{cb1}
and use ChPT in order to convert the result to the physical matrix
element.  ChPT thus ``corrects for" the 
systematic error from using unphysical (energy non-conserving)
momenta, and unphysical quark masses.  In addition, ChPT
can also be used to estimate effects from finite-volume,
excited states and (partial) quenching.  Whether this works
in practice depends on how well the chiral expansion converges
for the situation at hand.  For the
physical amplitude, one finds, for $\Delta I=3/2$ (in infinite
volume),\cite{bpp,golleuweak}
\begin{equation}
\langle\pi^+\pi^0|O_{\rm weak}|K^+\rangle_{\rm p}
=-\frac{4i\alpha^{27}}{f^3}(m_K^2-m_\pi^2)\left(\!1+{\rm chiral\ 
logs}(\Lambda)+d(\Lambda)\frac{m_K^2}{(4\pi f)^2}\right),\label{PHYS}
\end{equation}
where $\alpha^{27}$ is the $O(p^2)$ $\Delta I=3/2$ LEC, and
$d(\Lambda)$ is a linear combination of $O(p^4)$ LECs,
at the scale $\Lambda$ (we ignore an $O(p^4)$ term proportional to
$m_\pi^2$). $f$ is the pion-decay constant in the
chiral limit.  For physical pion and kaon masses, and at a
scale $\Lambda\sim m_\rho$ to $1$~GeV, the chiral logs are rather
small.  Phenomenological estimates of $\alpha^{27}$ and $d(\Lambda)$
exist.\cite{kametal}  For the unphysical matrix element, assuming
degenerate valence quark masses corresponding to a valence
meson mass $M_\pi$,
one finds\cite{golleuweak,golleu}
\begin{eqnarray}
\langle\pi^+\pi^0|O_{\rm weak}|K^+\rangle_{\rm u}
&=&-\left(\frac{4i\alpha^{27}}{f^3}\right)_{\rm PQ}2M_\pi^2
\left(1-N\frac{M_{VS}^2}{(4\pi f)^2}\left[
\log\frac{M_{VS}^2}{\Lambda^2}+d'(\Lambda)\right]\right.
\nonumber \\
&&\!\!\!\!\!\!\!\!\!\!\left.
+\frac{M_\pi^2}{(4\pi f)^2}\left[
-3\log\frac{M_\pi^2}{\Lambda^2}+d''(\Lambda)+
\frac{17.8}{M_\pi L}+\frac{12\pi^2}{(M_\pi L)^3}\right]\right)\,.
\label{UNPHYS}
\end{eqnarray}
Here $N$ is the number of sea quarks, $M_{VS}$ is the mass of
a meson made out of a valence and a sea quark, and $d'$, $d''$
are $O(p^4)$ LECs of the (partially) quenched theory.  Note that
$\alpha^{27}$, $f$ and other LECs are different for different
values of $N$, and only equal to those of the real world for
$N=3$.  This result also holds for the quenched case, $N=0$.

The finite-volume effects are power-like (we ignored 
exponentially small corrections), and potentially large.  
For example, with $f=160$~MeV, $M_\pi=500$~MeV and $M_\pi L=6$, this
correction is about $22$\% of tree level.
They arise from the ChPT one-loop diagram in which the final-state
pions of the weak decay vertex rescatter.  In hamiltonian
perturbation theory this corresponds to a contribution due to
intermediate pions with a momentum ${\vec k}=2\pi{\vec n}/L$, leading to
an energy denominator
\begin{eqnarray}
\frac{1}{L^3}\sum_{\vec k\ne 0}\frac{1}{\sqrt{M_\pi^2+{\vec k}^2}
-M_\pi}&=&\int\frac{d^3k}{(2\pi)^3}\frac{1}{\sqrt{M_\pi^2+{\vec k}^2}
-M_\pi}\label{FINVOL} \\
&&+\frac{c}{M_\pi L^3}(M_\pi L)^2\left(1+O\left(\frac{1}{(M_\pi L)^2}
\right)\right)\;,\nonumber
\end{eqnarray}
where $c$ is a numerical constant (for a more extensive
discussion, see Bernard and Golterman\cite{bgpp} and refs. therein).

The combined one-loop corrections in eq.~(\ref{UNPHYS}) are
rather large for typical lattice values of the parameters, and 
tend to increase the size of the matrix element.\cite{golleuweak,golleu}
The most recent numerical computation of this matrix element was
done by JLQCD,\cite{jlqcd} in the quenched approximation ({\it i.e.}
$N=0$), at an inverse lattice spacing of $2$~GeV.  
It was found that eq.~(\ref{UNPHYS}) describes the numerical
results reasonably well, in particular the volume dependence,
which has no scale ambiguities.
They obtained $d''(m_\rho)/(4\pi)^2\approx 0.025$ and 
$d''(1~{\rm GeV})/(4\pi)^2\approx 0.015$, with the difference
in agreement with eq.~(\ref{UNPHYS}). 

A variant of this method is to take, on the
lattice, $m_K=2m_\pi$.\cite{italians}  In this case the matrix
element with all mesons at rest conserves energy, and, in that
sense, is physical.  It is also closer to the real world than
the case with $m_K=m_\pi$, the choice made by JLQCD and in
earlier lattice simulations.  However, chiral corrections are
typically still large.\cite{golleu2}

\subsection{$\Delta I=1/2$}

The matrix element for $\Delta I=1/2$ non-leptonic kaon
decay constitutes a more difficult case for many reasons.
First, lattice computations are harder because of the
so-called ``eye" diagrams (containing quark 
propagators starting and ending at the same space-time point).
Also, unlike the $\Delta I=3/2$ case, mixing with the lower-dimension
operator $(m_s-m_d){\overline s}\gamma_5 d$ occurs.  This 
problem can be avoided by choosing degenerate (valence) quarks,\cite{cb1}  
or, in the case that $m_K=2m_\pi$, by using an improved 
action.\cite{italians}  In practice, we are therefore
limited to the computation of the matrix element with $m_K=m_\pi$
or $m_K=2m_\pi$ and all mesons at rest, if we can only 
determine the leading large-time behavior of eq.~(\ref{CORR}).  

This means that ChPT will be needed in order to extract the
relevant LECs from the unphysical
matrix element, in order to calculate the physical matrix element.
The question arises again which order in ChPT will be needed in
practice.  At one loop a problem occurs (for both $\Delta I=1/2,3/2$), 
as it turns out
that the $O(p^4)$ LECs needed for the physical matrix element
cannot be determined from the unphysical matrix element.
This can be seen as follows.  With arbitrary external four-momenta
$p_K=(im_K,{\vec 0})$, $p_{\pi_1}$ and $p_{\pi_2}$ with
$p_{\pi_1}^2=p_{\pi_2}^2=-m_\pi^2$, the contribution from
$O(p^4)$ operators is expected to be of the form
\begin{equation}
\frac{1}{(4\pi f)^2}[Am_K^2+Bm_\pi^2+2Cp_{\pi_1}p_{\pi_2}
+D(p_Kp_{\pi_1}+p_Kp_{\pi_2})]\times{\rm tree\ level}\;,
\label{POL}
\end{equation}
where $A$, $B$, $C$ and $D$ are independent (linear combinations of)
$O(p^4)$ LECs.
For the physical matrix element we have $p_K=p_{\pi_1}+p_{\pi_2}$,
and the polynomial in square brackets in eq.~(\ref{POL}) reduces to
$[(A-C-D)m_K^2+(B+2C)m_\pi^2]$.
For the unphysical matrix element we have $p_{\pi_1}=p_{\pi_2}=
(im_\pi,{\vec 0})$, and the polynomial becomes
$[(A+B-2C-2D)m^2_\pi]$ for $m_K=m_\pi$, and
$[(4A+B-2C-4D)m^2_\pi]$ for $m_K=2m_\pi$.  
In neither case do we obtain the
desired combinations of $O(p^4)$ LECs from the unphysical matrix
element.  In fact, in the mass-degenerate case, new, 
total-derivative $O(p^4)$ operators 
contribute because in this case energy is not conserved.  At
$O(p^2)$ this does not happen because of CPS symmetry (CP followed
by $d\leftrightarrow s$, $m_d\leftrightarrow 
m_s$),\cite{cb2} but CPS symmetry is not sufficient at $O(p^4)$.
All this shows that the unphysical matrix elements do not
contain enough information at $O(p^4)$.

Last but not least, there are new problems associated with 
(partial) quenching.  Even in the mass-degenerate case, the
$\Delta I=1/2$ operator couples to the $\eta'$, and therefore
the matrix element is afflicted by the double pole in the $\eta'$
two-point function (eqs.~(\ref{NEUPROP},\ref{QNEUPROP})).
(If $SU(3)$ is unbroken, the $\eta'$ does not couple 
to the $\Delta I=3/2$ operator.)  It turns out that enhanced
chiral logs are either absent\cite{golpal99} ($\Delta I=1/2$
at $m_K=m_\pi$) or numerically small\cite{golleu2} ($\Delta I=3/2$
at $m_K=2m_\pi$), at one loop in ChPT.  However, in the 
$\Delta I=1/2$ case, enhanced finite volume effects 
occur at one loop,\cite{golpal99}
giving rise to ``corrections" of order $L$ to the matrix element!
The double pole can occur on the internal lines of the meson
rescattering diagram, leading to terms proportional to
({\it cf.} eq.~(\ref{FINVOL}))
\begin{equation}
\frac{1}{L^3}\sum_{\vec k\ne 0}\frac{1}{\left(\sqrt{M_\pi^2+{\vec k}^2}
-M_\pi\right)^2}\;. \label{FINVOLDP}
\end{equation}
In finite volume, the smallest momentum has $|{\vec k}|=2\pi/L$,
leading to contributions of order $(1/L^3)\times L^4=L$.  This 
is another manifestation of the infrared diseases afflicting the
(partially) quenched theory.  In the quenched theory ($N=0$),
enhanced finite-volume effects are proportional to $m_0^2$
({\it cf.} eq.~(\ref{QNEUPROP})), while in the partially quenched case
($N\ne 0$) they are proportional to $M_{SS}^2-M_{VV}^2$
({\it cf.} eq.~(\ref{NEUPROP})).  (For a more detailed version of this
argument, see Bernard and Golterman.\cite{bgpp})

All this means that we may have to be more modest, at least
at present.  While we may not easily acquire sufficient information
to determine non-leptonic decay rates to $O(p^4)$, it would
be already interesting to determine the $O(p^2)$ weak LEC
$\alpha^{27}$ and the corresponding octet LEC $\alpha^8_1$
from the lattice.  This, however, can be accomplished using
simpler weak matrix elements, which avoid the enhanced finite-volume
complication described above.

\subsection{$K\to\pi\pi$ from $K\to\pi$ and $K\to 0$}

An easy way to explain this idea\cite{cb2} is by working at tree
level in ChPT first.  The physical matrix elements of interest
are
\begin{eqnarray}
\langle\pi\pi|O_{\rm octet}|K\rangle&=&\frac{4i}{f^3}(m_K^2-m_\pi^2)
\alpha^8_1\;,\label{KPIPI} \\
\langle\pi\pi|O_{\rm 27-plet}|K\rangle&=&-\frac{4i}{f^3}(m_K^2-m_\pi^2)
\alpha^{27}\;,\ \ \ (\Delta I=\frac{1}{2},\frac{3}{2})\;, \nonumber
\end{eqnarray}
where $\alpha^{27}$ and $\alpha^8_1$ are the $O(p^2)$ weak LECs.  However,
we can also consider the unphysical, but simpler matrix elements
\begin{eqnarray}
\langle\pi|O_{\rm octet}|K\rangle&=&\frac{4}{f^2}M^2(\alpha^8_1
-\alpha^8_2)\;, \label{KPI} \\
\langle\pi|O_{\rm 27-plet}|K\rangle&=&-\frac{4}{f^2}M^2\alpha^{27}\;,
\ \ \ (\Delta I=\frac{1}{2},\frac{3}{2})\;,\nonumber \\
\langle 0|O_{\rm octet}|K\rangle&=&\frac{4i}{f}(m_K^2-m_\pi^2)
\alpha^8_2\;,\nonumber
\end{eqnarray}
where in the $K\to\pi$ case we choose $m_K=m_\pi=M$. $\alpha^8_2$ is
the $O(p^2)$ LEC associated with the weak mass term.\cite{cb2,bpp,golpal}
There are no FSI, and therefore none of the
unphysical effects on the lattice associated with them.\footnote{Another
issue is whether ChPT at $O(p^4)$ takes {\it physical} FSI reliably
into account.  See Pallante and Pich\cite{fsi} and refs. therein.
Their results for FSI corrections to $\alpha^8_1$ appear to be
in reasonable agreement with the one-loop corrections of 
ChPT.\cite{kametal}}
One-loop corrections to eq.~(\ref{KPI}) have been calculated in
(P)QChPT including all contributions from $O(p^4)$,\cite{golpal,golpal99} 
and are typically found to be large (assuming a natural cutoff and
no sizable cancellations between chiral logs and $O(p^4)$ contact
terms).  A number of lattice collaborations are presently pursuing
this approach, and it should be interesting to see what can be learned
by fitting numerical results to $O(p^4)$ ChPT.  If, for certain
results from the lattice, the convergence of ChPT is reasonable,
one obtains the $O(p^2)$ LECs $\alpha^{27}$ and $\alpha^8_1$.  Even though
that does only give kaon decay rates to lowest order in ChPT, lattice
results for these LECs are interesting by themselves, and can be
compared with phenomenological estimates.\cite{kametal}  Of course
one would also obtain information about certain $O(p^4)$ LECs from
such fits, but, as in the case of unphysical $K\to\pi\pi$ matrix elements,
one does not get enough information to determine all $O(p^4)$ LECs
relevant for the {\it physical} matrix elements.\cite{golpal}  Results
for $\alpha^{27}$ and $\alpha^8_1$ from $K\to\pi$ and $K\to 0$ can
of course be checked against those obtained from
unphysical $K\to\pi\pi$ matrix elements, in principle.  

\section{Conclusion}

In this talk, I have argued that ChPT plays a crucial role in
extracting physics from numerical computations using Lattice QCD.
The generic setting of such computations is that of partially
quenched QCD, and ChPT can be systematically adapted to this
situation.  In fact, PQQCD with three light
quarks subsumes the real world, {\it i.e.} unquenched QCD with
valence-quark masses equal to those of the sea quarks.
It follows that LECs of the partially quenched theory are the
same as those of the real world.  This means that they can
often be determined from computations using unequal (and thus 
unphysical) values
of the sea- and valence-quark masses, allowing to pick choices
which are accessible with presently available computer resources
({\it cf.} section 2.1).  

In contrast, quenched QCD has no sea quarks, and is therefore 
different from QCD.
Because it is relatively cheap to simulate, it is a useful tool
for investigating the issues discussed in this talk, for instance
the convergence of ChPT at realistic values of the parameters.

In the second part, I addressed non-leptonic kaon decays.  
These involve the computation of weak matrix elements with more
than one strongly interacting particle in the final state,
and this leads to several difficulties.  The first difficulty
is that, when final-state interactions are present, one 
does not easily obtain matrix elements for a physical choice
of momenta.  This difficulty
originates in the fact that on the lattice we only have access
to the large-time regime of euclidean correlation functions.  The
second difficulty is that (partial) quenching artifacts, due to
the unphysical infrared behavior of (partially) quenched theories,
are more severe in this case.  However, ChPT provides an analytic
framework for parametrizing all these unphysical effects in terms
of the LECs of the physical theory (when three reasonably light
sea quarks are used on the lattice), and therefore can be used
to extract phenomenologically relevant information from
unphysical matrix elements.  In particular, I argued that it
should be possible to determine the leading weak LECs $\alpha^{27}$
and $\alpha^8_1$ reliably from Lattice QCD.

\section*{Acknowledgments}
I would like to thank the organizers for giving me the opportunity
to present this (selective) overview, and
Tomoteru Yoshie for making results from the CP-PACS collaboration
available before publication.  I thank Claude Bernard, Ka Chun Leung,
Elisabetta Pallante, Steve Sharpe, Noam Shoresh, as well as many
members of the CP-PACS collaboration for discussions.  I thank
Elisabetta Pallante and Steve Sharpe for comments on this manuscript.
This work is supported in part by the US Dept. of Energy.

\end{document}